\documentclass[amsmath,preprint,prd,
superscriptaddress,
nobibnotes,nofootinbib,floatfix]  
{revtex4}    
\usepackage{graphicx}
\usepackage{bm}

\def\ba{\begin{eqnarray}}
\def\ea{\end{eqnarray}}
\def\be{\begin{equation}}
\def\ee{\end{equation}}
\def\({\left(}
\def\){\right)}
\def\[{\left[}
\def\]{\right]}

\newcommand{\labeq}[1] {\label{eq:#1}}
\newcommand{\eqn}[1] {(\ref{eq:#1})}
\newcommand{\labfig}[1] {\label{fig:#1}}
\newcommand{\fig}[1] {Fig.~\ref{fig:#1}}

\begin{document}

\title{Cosmic Perturbations Through the Cyclic Ages}

\date{\today}

\author{Joel K. Erickson}
\email{jerickso@physics.columbia.edu}
\affiliation{ISCAP, Columbia Astrophysics Laboratory, New York, NY
  10027, USA} 
%\altaffiliation[Current Address: ]{ISCAP, Columbia Astrophysics
%  Laboratory, New York, NY 10027, USA}
%\affiliation{Joseph Henry Laboratories, Princeton University, 
%Princeton, NJ 08544, USA} 
\author{Steven Gratton}
\email{stg20@cam.ac.uk}
\affiliation{Institute of Astronomy, Madingley
 Road, Cambridge, CB3 0HA, UK} 
%\altaffiliation[Current Address: ]{Institute of Astronomy, Madingley
%  Road, Cambridge, CB3 0HA, UK} 
%\affiliation{DAMTP, CMS, Wilberforce Road, Cambridge, CB3 0WA, UK} 
\author{Paul J. Steinhardt}
\email{steinh@princeton.edu}
\affiliation{Joseph Henry Laboratories, Princeton University, 
Princeton, NJ 08544, USA} 
\author{Neil Turok} 
\email{N.G.Turok@damtp.cam.ac.uk}
\affiliation{DAMTP, CMS, Wilberforce Road, Cambridge, CB3 0WA, UK}

\begin{abstract}
We analyze the evolution of cosmological perturbations in the cyclic model, 
paying particular attention to their behavior and interplay  
over multiple cycles.  Our key results are: (1) galaxies and large scale 
structure present in one cycle are generated by  
the quantum fluctuations in the preceding cycle without interference from 
perturbations or structure generated in earlier cycles and without interfering
with structure generated in later cycles; 
(2) the  
ekpyrotic phase, an epoch of gentle contraction  with equation of state
$w\gg 1$ preceding the hot big bang, makes the universe homogeneous, 
isotropic and flat within any given observer's horizon; and, 
(3) although the universe is uniform within each observer's horizon, the 
global structure of the cyclic universe is more complex, owing to the 
effects of superhorizon length perturbations, and cannot be 
described in a uniform Friedmann-Robertson-Walker picture.  
In particular, we show that
the ekpyrotic phase is so effective in smoothing, flattening 
and isotropizing the universe within the horizon
that this phase alone suffices to solve
the horizon and flatness problems even without 
an extended period of dark energy domination (a kind of low energy
inflation).  Instead, the cyclic model rests on a genuinely novel,
non-inflationary mechanism (ekpyrotic contraction)
for resolving the classic cosmological conundrums. 

\end{abstract}

\maketitle

\section{Introduction}
\label{sec:introduction}

Cosmological observations support the idea that the part of
the universe we now observe emerged from a hot, radiation dominated and
expanding state. The original hot big bang theory did not specify
how this radiation was generated.
 Inflationary theory~\cite{Guth:1980zm} postulates a phase of
 superluminal expansion, driven 
by scalar field potential energy which ultimately decays into 
radiation. By contrast, in the
ekpyrotic~\cite{Khoury:2001wf} model, the radiation is 
generated by a brane collision, following an earlier empty phase.
The earlier phase is contracting from the viewpoint of Einstein-frame
four dimensional
effective theory. A transition from Einstein-frame contraction 
to expansion is also invoked in the pre-big bang model
 \cite{Gasperini:1992em}.

Both the inflationary~\cite{Bardeen:1983qw,Guth:1982ec,Hawking:1982cz} and 
ekpyrotic
models~\cite{Khoury:2001wf,Khoury:2001zk,Tolley:2003nx} plausibly give a 
spectrum of primordial fluctuations that is nearly scale invariant and
adiabatic, in
accordance with recent cosmic microwave
background~\cite{Spergel:2003cb} and large scale
structure~\cite{Tegmark:2003ud} observations.  Indeed, these two
theories seem to be the only ones that are generically able to give
such predictions~\cite{Gratton:2003pe}.

However, both theories are incomplete.  For inflation, it seems inevitable
that an initial singularity is still present~\cite{Borde:2001nh} (but
see~\cite{Aguirre:2001ks,Aguirre:2003ck}) only a 
finite affine parameter distance away, meaning that initial conditions
at the singularity are potentially significant.  In the ekpyrotic
scenario, the  initial
emptiness of the contracting phase is not explained.
   
The cyclic model~\cite{Steinhardt:2002ih,Steinhardt:2001st} builds on the ekpyrotic one,
essentially stacking 
a whole series of ekpyrotic histories together.  One post-bounce
expanding phase links onto another contracting phase. 
This contracting phase passes via a bounce into a new expanding phase,
and so on.  The universe evolves cyclically,
so that the puzzle of a ``fundamental beginning of time''
is at least deferred into the very distant past, 
and perhaps avoided altogether.

The ekpyrotic and cyclic models both have a higher
dimensional interpretation, inspired by M-theory.  This perspective is
critical for understanding how perturbations can propagate through a
bounce, and much recent work has focussed on this
issue~\cite{Tolley:2003nx,Craps:2003ai,Turok:2004gb}.  (For varying four 
dimensional
perspectives on the matching see
\cite{Khoury:2001zk,Durrer:2001qk,Durrer:2002jn,Cartier:2003jz,
Brandenberger:2001bs,Finelli:2001sr,Lyth:2001pf,Hwang:2001ga,
Peter:2002cn,Creminelli:2004jg}.)  Here we address other critical
issues for perturbations in the cyclic model, and are
able to use a four dimensional effective description almost
    throughout. We simply 
assume the essential features of the five dimensional matching
prescription are correct, and apply them to the four dimensional
effective theory.

Our analysis shows that
the galaxies and large scale 
structure in any given cycle can be generated by  
the quantum fluctuations in the preceding cycle without interference from 
perturbations or structure generated in earlier cycles and without interfering
with structure generated in later cycles.   
The global structure of the cyclic universe is more complex:
although the universe can be described as a nearly  homogeneous 
and isotropic
within any observer's horizon, 
the  global structure   
 cannot be characterized
by a uniform Friedmann-Robertson-Walker picture.
Our results further show that the ekpyrotic phase alone is 
sufficient for resolving the horizon and flatness problems and 
that an extended phase of dark energy domination or any other 
form of inflation is completely unnecessary.  This makes it
clear that the cyclic model is a genuinely novel, non-inflationary
approach to cosmology.

\section{Overview of the Cyclic Model and its Perturbations}
\label{sec:overview}

The cyclic model assumes that we live on a brane in a special
configuration of a higher dimensional theory such as M-theory.  Away
from a bounce, the  
universe can be treated using a four dimensional effective theory
consisting of   
gravity coupled to one or more scalar fields.
Assuming
the background universe is spatially flat, the metric is
\ba
ds^2=-dt^2+a^2 (t)\, \delta_{ij} dx^i dx^j
\labeq{frw}
\ea
where $a(t)$ is the scale factor. 
The main imprint of the higher dimensional theory on
the effective picture is through the addition of one or more scalar
fields $\phi$ with a potential $V(\phi)$.    
This potential performs many functions in the cyclic model, including
that of describing the dark energy responsible for the cosmic
acceleration observed today. Through most of this paper, we shall describe
the model in terms of a single scalar field, although we note that
generically more than one scalar field is involved. The scalar field 
$\phi$ satisfies 
\ba
\ddot{\phi} + 3 H
\dot{\phi} = - V_{,\phi},
\labeq{phieq}
\ea 
in the background \eqn{frw}, where dots denote
derivatives with respect to $t$ and $H \equiv \dot{a} /a$.  Ignoring,
for simplicity,
the coupling between ordinary matter and $\phi$, the Friedmann equation is
\ba H^2 = \frac{1}{3} \( \rho +
\frac{1}{2} \dot{\phi}^2 + V (\phi) \)
\labeq{heq}
\ea in reduced Planck units ($8 \pi G=1$), where $\rho$ is the
energy density of ordinary matter and radiation.

The potential $V(\phi)$ is chosen by hand at present, but should ultimately be
derivable from the higher dimensional theory.  It must be of a
certain form in order for the cyclic model to work \cite{Khoury:2003rt}. A 
useful
potential with the desired properties is
\ba
V(\phi)=V_0 \(e^{b\phi}-e^{-c\phi}\)F\(\phi\)
\labeq{vpot}
\ea
(see~\fig{potplot}).  Here $V_0$ is of order today's dark energy density,
$b$ is non-negative (and typically $\ll 1$) and $c$ is positive (and
typically $\gg 1$). $F(\phi)$ is a function whose precise form is
unimportant, but which tends to unity for $\phi$ greater than
$\phi_\text{end}$ and to zero for 
$\phi$ less than
$\phi_\text{end}$.
The resulting potential $V(\phi)$ has a large 
negative minimum, denoted $V_\text{end}$, at 
$\phi_\text{end}$. While the explicit exponential form used here 
is convenient for analysis, note that the cyclic model
actually works for a very wide range of potential forms, 
the only conditions being that they have a steep, negative 
and strongly negatively curved region over the observationally
relevant range of the scalar field. 
\begin{figure}
\includegraphics[width=14cm]{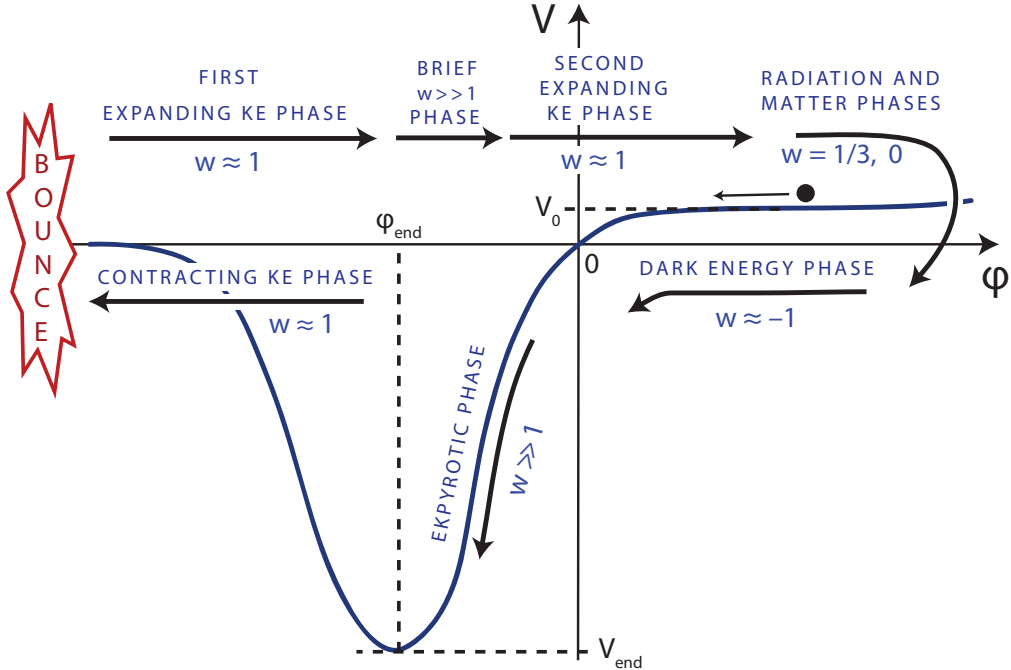}
\caption{
  \labfig{potplot} 
An example potential $V(\phi)$.  This plot shows
  where $\phi$ is on its potential at each stage in a cycle. The equation 
of state parameter of the background solution is denoted by $w$.
}
\end{figure}

Of central importance to the cyclic model is the {\it ekpyrotic } phase, 
in which the universe is slowly contracting and the 
scalar field is rolling slowly down its 
steeply declining, negative potential. For our example potential,
the negative exponential dominates, 
$V(\phi ) \approx -V_0 e^{-c \phi}$, and the background
universe enters an attractor scaling solution,
\ba
a(t) \propto (-t)^{2/c^2} \propto e^{\phi/c}, \qquad H={2\over c^2 t} 
\propto - e^{-{c \phi /2}}, \qquad w \approx c^2/3 \gg 1,
\label{scalingsol}
\ea
in which $t$ is negative and increasing, and $w$ is the ratio of the pressure to 
the energy density.  Notice that, since $c\gg 1$, as
the scalar field moves 
over a substantial range in Planck units towards $\phi_\text{end}$, 
the scale factor contracts 
by only a modest factor. In contrast, the Hubble parameter $H$ grows
dramatically, beginning from values comparable to today's value, 
and growing to values corresponding to  high energy scales. 

The scaling solution is only relevant as long as $\phi>\phi_\text{end}$,
and the function $F(\phi)$ is effectively unity. Once $\phi$ 
passes the potential minimum, the potential energy is quickly
converted to kinetic energy and the solution enters a kinetic energy 
dominated phase,
with 
\ba
a(t) \propto (-t)^{1\over 3} \propto e^{\phi/\sqrt{6}}, \qquad H={1\over 3 
t} \propto - e^{-{\sqrt{2/3}} \phi }, \qquad w \approx 1.
\labeq{kineticsol}
\ea
When lifted to higher dimensions, this solution describes two colliding 
branes (one with positive tension and the other with negative
tension), whose scale factors remain finite even as the four
dimensional 
scale factor $a(t)$ tends to zero and the scalar field $\phi$ tends to 
$-\infty$. Near the collision, the four-dimensional Einstein-frame 
metric and scalar field become singular coordinates; 
however, five-dimensional quantities like 
the metric on each brane, and the inter-brane distance, 
are perfectly finite. 
The matching of perturbations across
the bounce is therefore performed within the 
higher dimensional setting.

As the branes emerge from the collision, the solution followed is 
nearly the exact time-reverse of \eqn{kineticsol}; the
radiation and matter produced at the bang and a modest
enhancement of the kinetic energy of $\phi$ have a negligible 
effect while $\phi < \phi_\text{end}$.   There is a brief 
$w\gg 1$ expanding phase after $\phi$ passes $\phi_\text{end}$ 
moving to positive values, but the excess kinetic energy in $\phi$
quickly overwhelms the potential energy $V(\phi)$ and the universe
enters a second expanding kinetic phase (Figure 1). As is shown
in the Appendix, the expanding $w\gg 1$ phase is of modest duration
and for the remainder of this paper it can be safely ignored. 
It is then convenient to describe all three kinetic energy dominated 
phases, namely the contracting
and expanding kinetic phases with $\phi < \phi_\text{end}$, and
the second expanding kinetic phase with $\phi > \phi_\text{end}$,
as a single kinetic phase, and we shall generally adopt this 
terminology throughout the remainder of this paper. 

As we continue into the expanding phase, 
the kinetic energy in $\phi$ 
redshifts away as $a^{-6}$
and the universe becomes dominated by the radiation that was
produced at the bounce.  The net expansion in the entire
kinetic phase is $\sim e^{2\gamma/3 }$, where
$\gamma \equiv \ln
((-V_\text{end})^{1/4}/T_\text{rh})$, and $T_\text{rh}$ 
is by definition the temperature of the radiation when
it comes to dominate. As shown in 
Ref.~\cite{Khoury:2003rt}, cyclic models require $\gamma \sim 10-20$ 
in order to be compatible with observation. 
The additional 
Hubble damping due to the radiation
has the effect of slowing $\phi$ down to a halt on the
positive potential plateau. Then, the scalar field 
begins to gently roll downhill. The matter era
passes and the universe enters the dark energy phase.
Eventually, the rolling of $\phi$ carries it off the
plateau. The 
accelerated expansion due to dark energy 
reverses to slow ekpyrotic contraction.
The universe heads towards the next bounce and the
next cosmic cycle.

The
evolution of the background universe is illustrated by the inner,
solid, track
of the ``wheel'' of~\fig{pertwheel}. \fig{potplot} has also been labelled to
show where $\phi$ is on its potential at each stage in the cycle.  See
also~\fig{timeline} for a summary of the behavior of key
quantities. 

\begin{figure}
\includegraphics[width=12cm]{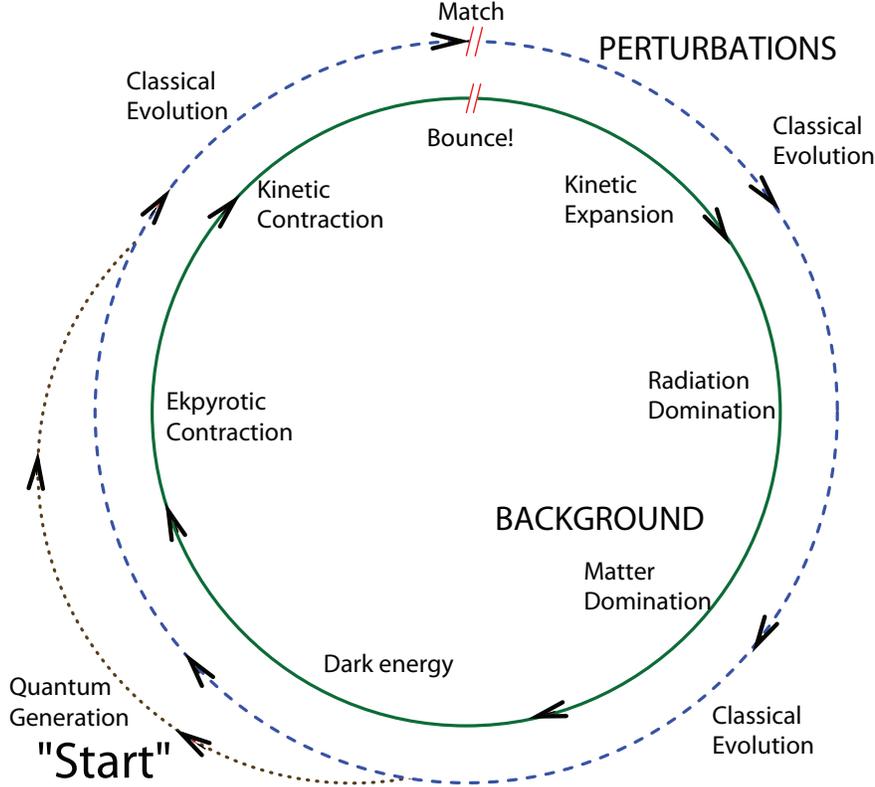}
\caption{
  \labfig{pertwheel} 
A ``wheel'' diagram indicating the behavior of both the background
  solution (inner, solid, line) and perturbations (outer, dashed and
  dotted, lines) in the cyclic model. ``Start'' marks the
  point in the cycle at which our perturbation analysis in
  Sec.~\ref{sec:quantum} begins.
}
\end{figure}
 
Note that while the bounce itself is nearly symmetrical, the 
background evolution for 
$\phi > \phi_\text{end}$ is highly asymmetrical, and 
the scale factor undergoes a large net
expansion from cycle to cycle. As explained above, the 
kinetic phase
gives a net expansion of ${2\over 3} \gamma $ e-folds.
The ensuing radiation phase gives a large number of e-folds
of expansion, and the matter phase adds a few more.  We may
approximate the combined number from the latter two phases as
$N_\text{rad}\equiv \ln (T_\text{rh}/T_0)$, where
$T_0$ is the cosmic microwave
background temperature today.  Dark energy adds another potentially
large number of e-folds $N_\text{dark}$.  By contrast, in the
ekpyrotic contraction phase, the scale factor contracts by
a very modest factor (from Eq.~\ref{scalingsol}, $a\propto H^{2/3 w}$). 
So there is a large net expansion every cycle 
of approximately
$2\gamma/3 + N_\text{rad}+ N_\text{dark}$ e-folds, which is
critical for the fate of the model when perturbations are considered
(see Sec.~\ref{sec:potential}). The large net expansion also plays a
key role in diluting the entropy density from cycle to cycle, and,
as we shall see in Sec.~\ref{sec:flatnesspuzzle}, in the cyclic model's
solution to the flatness, isotropy and horizon 
puzzles.

While the scale factor grows with each new cycle, 
locally measurable quantities like the Hubble parameter
and the density undergo periodic evolution. 
The Hubble parameter decreases by
$ 2 \gamma$ e-folds in the second kinetic energy dominated expanding
 phase, and by 
$2 N_\text{rad}$ e-folds in the ensuring radiation phase.
By contrast, in the ekpyrotic contracting phase,
$H$ increases in magnitude very rapidly, by a total of 
$N_\text{ekp}\equiv \ln\(\sqrt{-V_\text{end}/V_0} \)$ e-folds.
Since $V_0 \sim T_0^4$ in  order of magnitude, we
find $N_\text{ekp} \approx 2(N_\text{rad}+\gamma)$, which is 
also the
condition that 
the Hubble constant returns to its original magnitude
after a cycle. 

In this paper, for the study of perturbations, 
 we shall consider a universe that contains, in addition
to the scalar field $\phi$, cold dark matter and radiation, 
which we treat as
perfect fluids.  
This description captures
the broad features of cosmology well enough for our purposes.
The metric has scalar, vector and tensor
perturbations.  We  assume that the perturbations can be
well treated in linear theory, in which case the three sectors
decouple and we can follow perturbations Fourier mode by Fourier mode.
We ignore the vectors and tensors, and concentrate on the
scalar sector, in which the matter density perturbations exist.
We  work in
longitudinal gauge, in which there are no time-space or traceless
space-space perturbations.  Perfect fluids and scalar fields do not
support anisotropic stress, so the gravitational potentials are equal,
and the perturbed metric reduces to: 
\begin{equation} 
ds^2=-(1+2\Phi)dt^2+a^2(t)\,(1-2\Phi)\delta_{ij}dx^idx^j,
\labeq{perturbed}
\end{equation}
where $\Phi$ is the Newtonian potential.

Within any one cycle, the Einstein and matter equations fully
determine the classical evolution of the perturbations.  This
classical evolution is indicated by the outer, dashed loop of our wheel diagram
(\fig{pertwheel}).

The cyclic model, like inflation, relies on the amplification of
quantum fluctuations to initiate structure formation.  It is helpful
to think in the Heisenberg picture.  Here the quantum field mode
operators satisfy the classical equations of motion but even if the
quantum expectation value of a mode amplitude is zero, the
expected variance cannot also be zero (just as for the ground state of
a simple harmonic oscillator, for example).  When the equation 
governing a mode moves from having oscillatory solutions to growing
    and decaying
solutions, 
%NNN the mode can start growing and then 
the quantum
variance will also grow, as the square of the classical growing mode
amplitude. As far as the evaluation of
future expectation values is concerned, it now becomes possible to
accurately approximate the quantum picture with 
a classical one in which the mode amplitude is treated as
a random variable with  mean and variance given by the 
quantum calculation. We  say that a
perturbation has been \textit{generated} when the classical
probabilistic
description becomes accurate.  In the cyclic model perturbations can be
generated during both the dark energy and ekpyrotic phases,
    and  quantum fluctuations in one cycle  become classical
    stochastic perturbations by the next.  Of
course, this stochastic contribution to a mode amplitude is only
important if it is comparable to or greater than that which is already
there from 
the classical evolution.  ``Quantum generation'' of perturbations is represented 
by the dotted line in the wheel diagram, \fig{pertwheel}.

To  complete the perturbation loop in our
wheel diagram we  need to know how to match perturbations
across a bounce. This is the one place where
the four-dimensional effective picture becomes invalid and results obtained in 
higher dimensions must be used.
Recent work~\cite{Tolley:2003nx,Turok:2004gb,McFadden:2005mq} suggests
how this occurs, with 
long-wavelength growing modes going in to the crunch matching onto
growing modes going out from the bang. This matching
occurs in a manner that, for long wavelengths, is independent of wavelength.

Hence we are now able to follow perturbations through
multiple cycles in the cyclic model, allowing for both quantum
generation and matching across the bounce in addition to the classical
evolution.

\section{Quantum Generation of Perturbations}
\label{sec:quantum}

In principle, when describing a cyclic model, one can start anywhere
on the wheel diagram, \fig{pertwheel}. For simplicity, we
start well into a long-lasting
dark energy phase in which the universe has become very 
homogeneous and flat and any pre-existing matter, radiation,
or scalar field 
perturbations have been redshifted away to negligible levels.
Later on, in Sec.~\ref{sec:potential}, we show that
%it is not necessary for the consistency of the model to assume 
%a long-lasting dark energy phase.
the consistency of the model does not require that the dark energy
phase be long-lasting.

In a universe containing only a scalar field $\phi$, the Einstein equations fix
the scalar field fluctuation 
$\delta \phi$ in terms of the Newtonian potential $\Phi$ and its time
derivative:
\ba
\frac{\dot{\phi}}{2} \delta \phi = \dot{\Phi}+H \Phi.
\labeq{link}
\ea
 Thus there is only one true scalar degree
of freedom, which we take to be $\Phi$, and $\Phi$ satisfies the
second-order differential equation: 
\begin{equation}
\ddot{\Phi} + \biggl( H - \frac{2\ddot{\phi}}{\dot{\phi}} \biggr) \dot{\Phi} +
2  \biggl( \dot{H} - \frac{H \ddot{\phi}}{\dot{\phi}} \biggr) \Phi -
\frac{\nabla^2 \Phi}{a^2} =0.
\labeq{newteq}
\end{equation}
This equation can be used both for the classical evolution of $\Phi$
and, as discussed above, for determining the variance of the
fluctuations generated quantum mechanically.  Since, by assumption,
there are initially no classical perturbations, we turn to the quantum case.

For the quantum calculation we need to pick a suitable quantum state
for each Fourier mode.  Just as in inflation, we assume that when 
the evolution of a perturbation mode is ``gradient dominated'' (sometimes
    called ``subhorizon''), the  scalar field fluctuation  
$\delta \phi$ is in the appropriate incoming adiabatic vacuum state. 
Eq.~\eqn{link} is then used to determine the
state of $\Phi$ in this period  (see e.g.~\cite{Khoury:2001zk}).
We then evolve forward in time using~\eqn{newteq} until the 
spatial gradients become negligible in the time evolution. Now, the
mode is said to evolve in an ``ultralocal'' (sometimes called
    ``superhorizon'') manner.  
The modulus squared of the mode amplitude then gives the
quantum variance which, when the quantum picture is replaced
by the stochastic classical one, becomes the
classical power spectrum on that scale.  Repeating the calculation for
different 
comoving wavenumbers allows us to build up the complete power spectrum.  The
power on a given scale changes with time in
accordance with Eq.~\eqn{newteq}, but all modes that are
in the long-wavelength, ultralocal  regime will evolve 
in concert.

We perform the above procedure, solving Eq.~\eqn{newteq} mode by mode
with the appropriate initial conditions, to build up a power spectrum
for the perturbations.  All modes of interest start off gradient dominated
and end up in the ultralocal  regime. Longer wavelength
modes start to follow ultralocal evolution sooner, shorter wavelength
modes later. The 
very shortest wavelength modes go ultralocal only in the kinetic energy 
dominated
phase just before the big crunch. Once all modes have gone ultralocal,
the whole power spectrum evolves in concert and simply grows in
amplitude as the bounce is approached. 

The detailed
shape of the power spectrum depends on the exact background evolution
and the specific details of the scalar field
potential $V(\phi)$. 
\fig{primpower} shows a power spectrum
for a typical model (with $b=0.1$ and $c=30$), evaluated close to the
crunch when all the modes are evolving ultralocally. 
The comoving wavenumber
is denoted by $k$.  Observe that
there are large bands of $k$ for which the power spectrum is almost
scale invariant.  Note that there is a
feature in the power 
spectrum on scales $k_\text{tran}$ that went ultralocal at around the
time of the 
transition from expansion to contraction.  Also note that modes on
larger scales (which went ultralocal in the dark energy phase) have
a comparable amplitude to those on smaller scales (which went
ultralocal in the ekpyrotic contraction phase).

%\footnote{This might at
%first sight come as a surprise to those accustomed to inflationary
%perturbation theory (at least it did to us): in an expanding universe,
%the comoving curvature perturbation, and so roughly $\Phi$, is
%generated at an amplitude given by the inflationary energy scale and is then
%conserved whilst the mode is superhorizon or long-wavelength.  The same logic
%applied here would lead one to expect that the large-scale
%perturbations should have an amplitude given by the dark energy scale,
%exceedingly small.  The resolution of the argument is that the
%approximate constancy of $\Phi$ on superhorizon scales does not carry
%over from an expanding universe into a contracting universe; even
%long-wavelength perturbations get amplified during contraction.}.

\begin{figure}
\includegraphics[width=12cm]{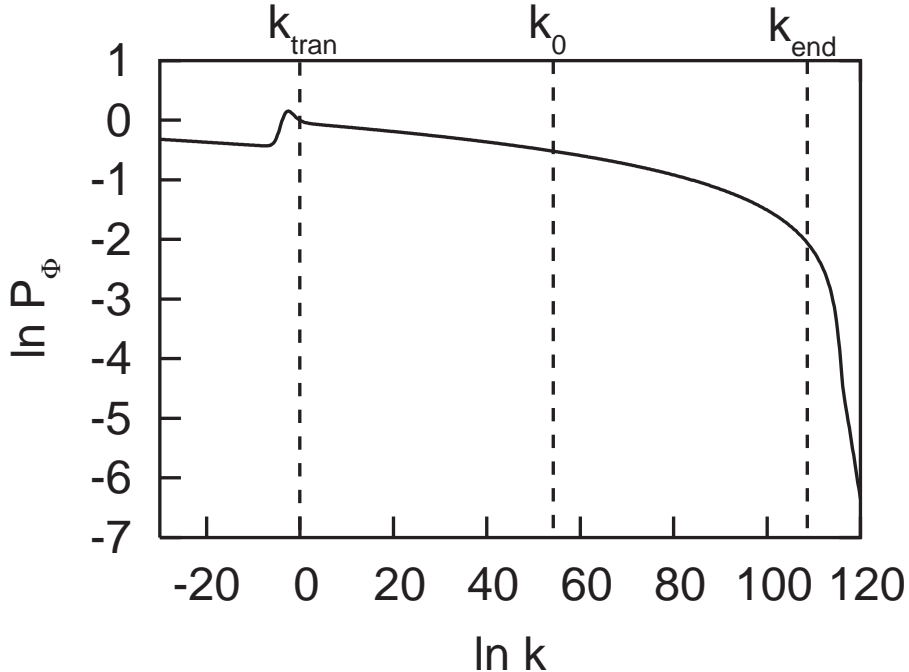}
\caption{
  \labfig{primpower} 
A plot of the power spectrum $P_\Phi$ of quantum-generated fluctuations in the
  Newtonian potential $\Phi$ going
  into the crunch.  A horizontal line corresponds to scale
  invariance.  $k_\text{end}$ indicates the wavenumber of the last
  modes to go ultralocal during ekpyrotic contraction.  $k_0$
  indicates the scale corresponding to our horizon today. The parameters chosen
  in Eq. \eqn{vpot} are $b=0.1$ and $c=30$.
}
\end{figure}
 
The range in $\ln k$ of modes to
the left of the feature   
is of order $N_\text{dark}$, the number of e-folds of dark energy
expansion, while the range of modes to the right of the feature 
is of order $N_\text{ekp} \equiv \ln \sqrt{-V_\text{end}/V_0}$.  
As we  explain later (Sec.~\ref{sec:classical}), the value of $k$ for modes 
on our current
horizon scale
is approximately $ N_\text{ekp}/2$ e-folds
to the left of $k_\text{end}$ and hence near the middle of the
approximately scale-invariant region of the power
spectrum. 

\section{Matching Perturbations through the Bounce}
\label{sec:matching}

Now we  map perturbations through the singularity.  
First we  relate physical length scales on either side of the bounce.
Second, we relate the Newtonian potential for each mode on
either side of the bounce.

The first task is simplified by the near symmetry of the
contracting and expanding kinetic phases.  Consider the wavelength of
the last mode to go ultralocal during ekpyrotic contraction, with
wavenumber $k_\text{end}$ in \fig{primpower}.  Its
physical wavelength is roughly given by the Hubble radius at that time,
$\sim 1/ \sqrt{-V_\text{end}}$.  By the symmetry of the kinetic phases,
  its physical wavelength when $\phi$ reaches $\phi_\text{end}$ on the
  way out after the bounce will again be $\sim 1/
  \sqrt{-V_\text{end}}$.

For the second task, we need to know how the
Newtonian potential behaves on either side of the bounce in the
four-dimensional effective treatment.  All modes become
ultralocal as the bounce is approached, and so all behave in the 
same manner, independent of their wavelength.  If $t=0$ corresponds to the
bounce, the Newtonian potential goes like
$A+B \, t^{-4/3}$ on the way in and as
$A'+B' \, t^{-4/3}$ on the way out (see Sec. \ref{sec:global}).  So each side 
has a diverging term and a constant term.  On the way
in, the constant term is the ``decaying'' mode while the $t^{-4/3}$
term is the growing mode.  On the way out, the roles are reversed with
the constant term now the ``growing'' mode and the $t^{-4/3}$ term now
the decaying mode.

The perturbations are in the growing mode
approaching the bounce.  It is essential for the success of the cyclic
model that such perturbations lead to some growing-mode perturbations
after the bounce.  For the shape of the spectrum to be preserved it is
also essential that such mapping occurs in a manner which is at least
approximately independent of
wavelength.   

To perform the mode matching one must move from the four dimensional effective
theory into five dimensions (where the bounce now corresponds to the extra
dimension momentarily contracting to zero size and then expanding
again).  The work of Tolley, Turok and Steinhardt~\cite{Tolley:2003nx}
does this and provides us with a matrix $\bm{M}_\text{TTS}$ that
relates the incoming and outgoing mode coefficients.  
The form of this matrix is presented in
Sec.~\ref{sec:global}, after we have introduced normalized
mode functions.  
According to the matching prescription of \cite{Tolley:2003nx}, an
incoming growing mode maps onto an 
outgoing solution with a non-zero growing mode component. Furthermore,
the matching is independent of wavelength, as
required.  The procedure is to find a quantity that behaves as a
massless scalar field in a particular five dimensional gauge near the bounce.
Earlier work of Tolley and Turok~\cite{Tolley:2002cv} showed that
there is a natural way to analytically continue such fields through
the bounce.
By understanding the
correspondence between four and five dimensional perturbations one can thus 
match four dimensional
modes across the bounce.  Note that alternative matching prescriptions
with alternative matching matrices are straightforward to use in
place of $\bm{M}_\text{TTS}$.  (Indeed, very
recent work of McFadden, Turok and 
  Steinhardt~\cite{McFadden:2005mq} takes a wider five dimensional view of the
vicinity of the collision, and effectively pre- and post-multiplies
$\bm{M}_\text{TTS}$ by 
  another matrix.  However this does not alter the qualitative
  features of the matching, so $\bm{M}_\text{TTS}$ is used in this paper.)

The result of the matching is that after the bounce the ``primordial'' power 
spectrum for the expanding phase 
is in the ``growing'' mode and 
has just the same shape as the power spectrum before the bounce
(i.e.\ \fig{primpower} again).  The amplitude of the power spectrum of
the Newtonian potential $\Phi$ is
now roughly 
constant and is determined by three things: the starting amplitude,
given by the adiabatic vacuum assumption; the amount of growth
occurring during contraction; and finally the precise matching
coefficient of growing mode to growing mode in the matching matrix.
The parameters controlling the length of the contraction phase
(e.g. $\phi_\text{end}$) and the details of the bounce (such as the
relative speed of the branes at collision) must be chosen to make
the perturbation amplitude approximately $10^{-5}$, in order to match
observations.  

\section{Classical Evolution of Perturbations}
\label{sec:classical}

The ``primordial'' power spectrum computed at the beginning of
an expansion cycle can be evolved
straightforwardly  through to the end of 
the
matter epoch in order to compare it with observation.  
In this section, we track the perturbations
further around the wheel diagram into the dark energy and contraction
phases, in order to see what effect they 
have on the quantum generation of the next round of perturbations. 
We  solve the perturbed Einstein, fluid and scalar field
equations numerically mode by mode, starting deep within the radiation
era.  (All modes of interest follow ultralocal evolution in the
kinetic era, so we need not worry about the evolution there.)  Our
numerical code employs  
synchronous gauge, but we express the results in terms of the fully gauge-fixed
Newtonian gauge potential $\Phi$.

\begin{figure}
\includegraphics[width=16cm]{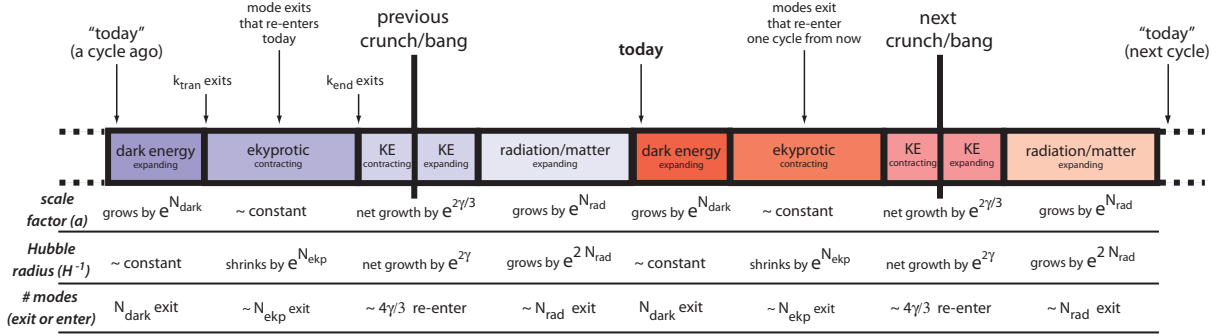}
\caption{
  \labfig{timeline} 
A timeline of the cyclic universe  showing the behavior of
 key quantities over the course of a cycle.
 The labels ``$k_\text{tran}$ exits'' and ``$k_\text{end}$ exits'' show the times
 where these two modes start to follow ultralocal evolution (i.e. when spatial gradients 
 become negligible).
}
\end{figure}

The perturbations are initially set in their adiabatic growing mode
when they are all ultralocal (i.e.\ outside the effective horizon).
Observations of the cosmic microwave background and large scale
structure are sensitive to scales from our cosmological horizon down
to roughly ten e-folds in $k$ smaller.  To relate this to our
simulations, we need to know what portion of the ``primordial''
spectrum is relevant to observation.  To do this, we work out the
difference in $k$ between modes on our horizon today and the last
modes generated during ekpyrotic contraction.  As mentioned in the
previous section the physical wavelength of the latter modes is
roughly $1/ \sqrt{-V_\text{end}}$ when $\phi$ passes $\phi_\text{end}$
in the expanding phase.  Since then there has been a brief $w\gg1$ phase,
a second kinetic energy dominated phase, and the 
radiation and matter phases (see \fig{potplot}), 
providing an expansion of
$e^{2\gamma/3+N_\text{rad}}$ between them.  So these modes have a
physical wavelength 
today of order $ e^{2\gamma/3+N_\text{rad}} /\sqrt{-V_\text{end}}$.
Modes on the horizon today (having a physical wavelength of order
$1/\sqrt{V_0}$) thus have a wavelength
$e^{-2\gamma/3-N_\text{rad}+N_\text{ekp}}$ times this, where
$N_\text{ekp}$ is as introduced previously in Sec.~\ref{sec:quantum}.
The universe reheats to a moderate temperature after the bang.  The
reheat temperature $T_\text{rh}$ is tuned to produce density
perturbations of the observed amplitude and to satisfy other
constraints.  This imposes a constraint on
$\gamma$, as defined in Sec.~\ref{sec:overview} via the relation   
$T_\text{rh}=e^{-\gamma} (-
V_\text{end})^{1/4}$ 
(which implies  $N_\text{rad}= -\gamma + N_\text{ekp}/2$):
namely,
one needs $\gamma \sim 10-20$ (or $e^{-\gamma}
\sim  10^{-(4-8)}$), depending on the value of $c$.  
For a more precise discussion
of the permitted range for $\gamma$, see Ref.~\cite{Khoury:2003rt}.  Thus, the 
modes on the horizon today
have a wavelength that is $ e^{-2\gamma/3-N_\text{rad}}
\sqrt{-V_\text{end}/V_0}=e^{\gamma/3+N_\text{ekp}/2}$ times that of
the last modes to be generated in the ekpyrotic phase; the former are a
factor of $\gamma/3+N_\text{ekp}/2 \gtrsim 5+N_\text{ekp}/2$ lower in
$\ln k$ than the latter.  Since $N_\text{ekp}$ is very large, of order
$100$, modes on 
our horizon today lie roughly in the middle of the logarithmic $k$
range of modes to the right of the transition feature seen
in~\fig{primpower}.  The power spectrum is very smooth in the relevant
10 e-folds in $k$ around this point, with only a slight tilt.  
In Table~\ref{tab:scales}, 
we
present a table of some of the scales mentioned in this paper 
and give a timeline of the model
in~\fig{timeline}.

\begin{table}
\caption{\label{tab:scales} A table showing various scales discussed
  in this paper 
  relative to today's horizon.  Note that $\gamma \equiv \ln
((-V_\text{end})^{1/4}/T_\text{rh})$, 
   $N_\text{ekp} \equiv \ln 
\sqrt{-V_\text{end}/V_0}$,
and $N_\text{rad}\equiv \ln
  (T_\text{rh}/T_0) = -\gamma + N_\text{ekp}/2$.}
\begin{ruledtabular}
\begin{tabular}{l|lc}
Length Scale & Size Relative to Today's Horizon \\
\hline
Today's horizon & 1 \\
Current wavelength of the modes that: & \\
\ldots were on the horizon one cycle ago & 
$e^{N_\text{dark}+ 2\gamma/3+N_\text{rad}}$ \\
\ldots will be on the horizon one cycle from now  & 
$e^{-N_\text{dark}-2\gamma/3-N_\text{rad}}$ \\ 
\ldots were the first ones to go ultralocal during & 
$e^{2\gamma/3+N_\text{rad}}$\\
~~~~~~~~the ekpyrotic phase one cycle ago & \\  
\ldots were the last ones to go ultralocal during & 
$e^{-N_\text{ekp}+2\gamma/3+N_\text{rad}}$\\
~~~~~~~~the ekpyrotic phase
one cycle ago & \\  
\ldots will be the first ones to go ultralocal during & 
$e^{-N_\text{dark}}$\\
~~~~~~~~the coming ekpyrotic phase of this cycle & \\  
\ldots will be the last ones to go ultralocal during&
$e^{-N_\text{dark}-N_\text{ekp}}$\\
~~~~~~~~the coming ekpyrotic phase of this cycle& \\  
\end{tabular}
\end{ruledtabular}
\end{table}

As an example displaying the qualitative behavior of the Newtonian potential,
we have studied the case for a potential of the form given
in~\eqn{vpot} with $b=0.1$ and $c=5$.
 The code
starts with seven
e-folds of radiation domination remaining.  This is followed by
seven e-folds of matter domination 
and then  only  two e-folds worth of dark
energy domination before ekpyrotic contraction begins. These parameters
are not far from those that might give  a fully realistic description of the
universe, and serve to illustrate the  qualitative features
of the mode evolution without requiring us to introduce  large exponential 
factors that complicate the numerics.

\fig{multipower} presents the Newtonian potential at four different times.  Note 
that, in this plot,
a horizontal 
line corresponds to scale invariance.  
Moreover, the tilt of the input power spectrum has been neglected, and thus 
\fig{multipower} depicts the transfer 
function for an exactly scale-invariant power spectrum of the
Newtonian potential.

\begin{figure}
\includegraphics[width=12cm]{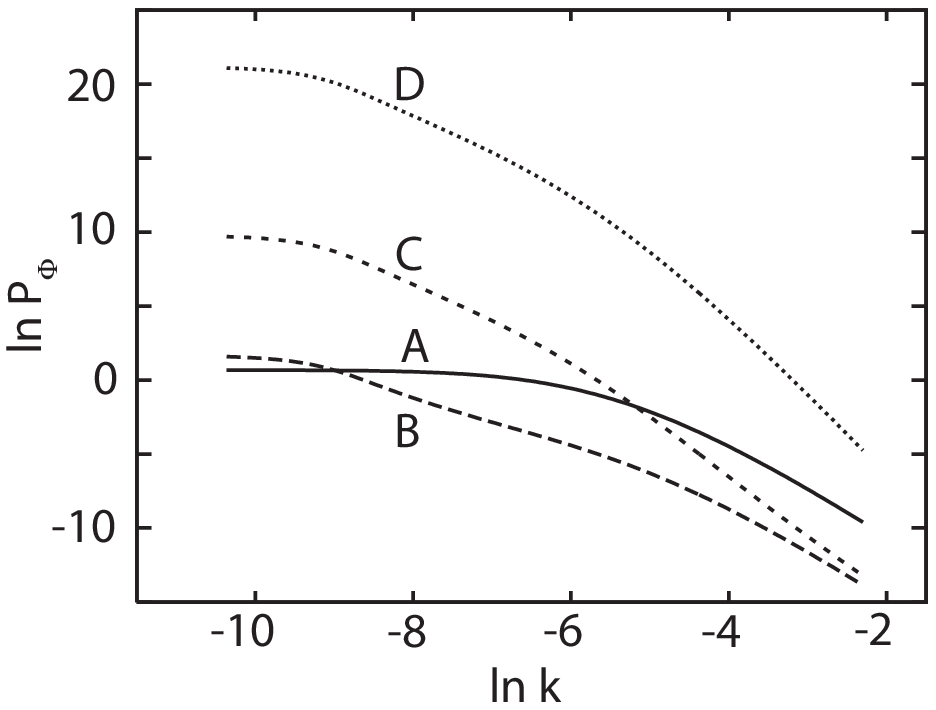}
\caption{
  \labfig{multipower} 
A plot of the power spectrum $P_\Phi$ of the Newtonian
 potential $\Phi$ (in arbitrary
  units) at various
  points in a cosmic cycle.  $k=1$ corresponds to the horizon at the
  start of the simulation in the radiation era, and $\ln k \approx
  -6.6$ corresponds to the horizon at matter-radiation equality.  A
  horizontal line 
  corresponds to scale 
  invariance.  The curve A indicates the power at the end
  of the matter epoch.  Curve B indicates the power at turnaround
  ($H=0$).  Curve C indicates the power some way into ekpyrotic
  contraction, and curve D indicates the power further into the
  ekpyrotic phase.  
}
\end{figure}

 In curve A, the power spectrum $P_{\Phi}$ is shown at a time corresponding to
``today'', the end 
of the matter epoch and the beginning of the dark energy epoch.  
The spectrum is scale 
invariant at small $k$ and has the expected bend at a scale corresponding to 
the horizon at matter-radiation equality. At larger wavenumber, the curve falls 
off
as $k^{-4}$.  
%The scalar field $\phi$ is nearly static, so 
The scalar field fluctuations
have had no 
significant effect on $P_{\Phi}$ at this stage. 
This curve is in perfect accord 
with observations.

%Note the usual bend in the spectrum at 
%a scale corresponding to the horizon at matter-radiation equality.  On
%small scales the curve behaves as $k^{-4}$ 
%(compare it to the $k^{-4}$
%guide line)
%.  The presence of the scalar field has so far made little
%impression on the Newtonian potential.  We thus have the key result
%that the cyclic model can successfully account for the formation of
%cosmic structure that we see.

Curve B shows the power spectrum evolved through the dark energy phase to 
turnaround at $H=0$.
On small scales the power has dropped uniformly as perturbations are
diluted.

Curves C and then D show the power spectrum at two stages in the
ekpyrotic contraction phase.  On large scales the spectrum grows rapidly, and on 
small scales
it reddens. 
The unstable scalar field contribution dominates
the Newtonian potential in this epoch, as matter and radiation are negligible.

\section{Interference of perturbations from different cycles}
\label{sec:potential}

We are now in a position to investigate whether
the current perturbations interfere with the quantum generation of the new.  
First, we
compare the amplitude of the current perturbations
with that of the new ones to be generated, neglecting any effect
that preexisting perturbations might have on the quantum generation of new 
perturbations.  Then we 
check if this approximation is justified. 

We  concentrate on scales that are observationally relevant
at the beginning of dark energy domination in the next cycle.  
The mode whose wavelength will equal the current 
horizon radius one cycle from 
now is today on a scale 
that is a factor of roughly $e^{-(N_\text{dark} + N_\text{rad}+2 \gamma/3)}$ 
times the
current horizon radius, 
  the inverse
of the total expansion of 
the universe in the interim, corresponding to 
$N_\text{dark}+N_\text{rad}+ 2 \gamma/3$ e-folds 
in $k$.  We saw above that the smallest wavelength mode produced in
the ekpyrotic phase lay roughly $N_\text{rad}+4 \gamma/3$ e-folds within the current
horizon.  So if there is a significant dark energy phase then there
will be negligible
power  on the scale of the horizon radius in the next
    cycle.  However, as we shall see, an extended period of dark energy
domination is not necessary for the cyclic model.  To prove the point,
we will consider the ``worst case scenario," that
the number of e-folds of dark energy domination is negligible.
This would put the horizon radius a cycle from now  near the high-$k$
downward turn of the ``primordial'' power spectrum.  Since we have
only worked to logarithmic accuracy in determining scales, let us
conservatively assume that the ``primordial'' spectrum is
still
scale invariant even on these small scales. 
Including the red tilt will only strengthen the argument
below showing that there is no obstruction to cycling due to build up of
perturbations from cycle to cycle.

We  estimate the amplitude of the current perturbations on the
relevant scale as follows.  The amplitude a mode on our
horizon scale has now is roughly $10^{-5}$.  In the
next section we show that during ekpyrotic contraction the mode grows like $1/t$
with $t$ here the time to the forthcoming bounce.  
From our numerical results we see that the tilt of the power spectrum
on subhorizon scales is reddening as we proceed to the bounce.
Asymptotically we expect the mode 
amplitude to drop as $k^{-4}$ based on the following argument: $\Phi$
  from the matter is two powers down from scale-invariant, and this
  sources a perturbation  $\delta \phi$ a further two powers down
  (see Eq.~\eqn{sfeq} below, neglecting the time derivative terms).
  Then when the scalar field is again dominant, Eq.~\eqn{link} tells
  us that $\Phi$ should now go like $\delta
  \phi$, four powers down.

Thus with $k_0$ corresponding to our horizon,
the amplitude of the current perturbations on a scale $k$ at a time
$t$ before the crunch is roughly 
\ba 10^{-5} \, \frac{t_0}{t}
\(\frac{k}{k_0}\)^{-4}, 
\ea 
where $t_0$ is the time from the start of
the ekpyrotic phase to the crunch.  So the amplitude on our future
observer's horizon is roughly 
\ba 10^{-5} \, \frac{t_0}{t} e^{-4N_\text{rad}}.  
\labeq{currentmode} 
\ea

Now let us estimate the amplitude of the new perturbation mode
generated on the same scale.  The quantum perturbations generated are
almost scale invariant, and we exploit this by actually calculating
the amplitude of a mode generated in the dark energy phase. The last
of these will have been generated just before $t_0$, the onset of
ekpyrotic contraction. Just as in slow-roll inflation, $\Phi$ will from
Eq.~\eqn{link} then have an amplitude of roughly $\dot{\phi} / H$
times the amplitude of the fluctuation in $\delta \phi$, which
is $H$.  So $\Phi$ is roughly $\dot{\phi}$ before turnaround, which is of order 
$H_0$, the Hubble constant today.  During the contraction the new mode
also grows like $1/t$.  So at a time $t$ before the crunch, the
newly-generated mode amplitude will be of order 
\ba 
H_0 \frac{t_0}{t}.
\labeq{newmode} 
\ea 
Comparing \eqn{currentmode} and \eqn{newmode} we
see that the time dependence cancels and we need the inequality \ba
10^{-5} e^{- 4 N_\text{rad}} < H_0 \ea to be satisfied in order for
the new perturbations to dominate over the old.

Rewriting $e^{N_\text{rad}}$ as $T_\text{rh}/T_0$ and $H_0$ as
$T_0^2$, we obtain a lower bound on $T_\text{rh}$:
\ba
T_\text{rh} > 10^{-5/4} \sqrt{T_0},
\ea
in reduced Planck units.  Thus the reheating temperature need only be
a few hundred 
GeV.  

As shown in Ref.~\cite{Khoury:2003rt}, this is not a difficult condition to
satisfy.  In that reference, it is shown that it is possible to have
perturbations with an amplitude of $10^{-5}$ and satisfy all other known
constraints for a wide span of $T_\text{rh}$ above a few hundred GeV ranging up
to $10^{10}$~GeV (or more, depending on the value of $c$).

So the lack of growth of modes that enter the horizon during the
radiation era, leading to a $k^{-4}$ drop in power on small scales,
combined with a further $k^{-4}$ drop during asymptotic scalar-field
domination, seems easily sufficient to ensure that new perturbations
will dominate over the current ones for an observer in the next cycle.
We have not even had to consider other effects that further suppress
small-scale power, such as dark matter free streaming, in order to
reach 
this conclusion.

We now check that current perturbations do not more
subtly influence the form of new ones by interfering with their
quantum generation.  Perturbations in the scalar field give the main
contribution to the Newtonian potential during ekpyrotic contraction,
since it is the scalar field that is dominating the matter
content of the universe during this phase.  The scalar field
perturbations satisfy: 
\ba
\ddot{\delta \phi} + 3 H \dot{\delta\phi} -\frac{\nabla^2 \delta
  \phi}{a^2} = - V_{,\phi\phi} \delta \phi +4 \dot{\Phi} \dot{\phi} -
2 \Phi V_{,\phi}.
\labeq{sfeq}
\ea
The $\Phi$ terms on the right hand side of \eqn{sfeq} provide the
opportunity for current perturbations in the Newtonian potential to
influence the generation of the new ones.   We need to check that
their contributions are much less than that of the $V_{,\phi\phi}
\delta \phi$ 
driving term.  A quick way to do this is as
follows.  We shall see in 
the next section that in a growing
ultralocal 
mode $\Phi$ goes like $H/a$.  We then use
Eq.~\eqn{link} and the background equations to deduce that $\delta
\phi$ behaves like 
$-\dot{\phi}/a$.  In the background solution both $H$ and $\dot{\phi}$
go like $1/t$, so $\delta \phi$ in Planck units is roughly
equal to $\Phi$.  Hence we need to compare $V_{,\phi \phi}$ times the
newly generated 
$\Phi$ to $V_{,\phi}$ times the pre-existing $\Phi$.  Now $V_{,\phi \phi}$ and
$V_{,\phi}$ are comparable in Planck units, and we have already seen
that the newly generated 
$\Phi$ is much larger than the pre-existing $\Phi$ on the scales of
interest.  Thus the influence of current perturbations on the
generation of the new perturbations is indeed negligible.

One might be concerned about the effect of nonlinearities in the
matter power spectrum on small scales in this discussion.  We do not
think that this is important however, because, even if the matter does go
nonlinear, this does not change the typical Newtonian potential very
dramatically.   
Furthermore, the scalar field does not couple effectively to the matter,
only via gravity.  
So both the Newtonian potential and scalar
field perturbations can be well approximated by linear perturbation
theory. 

Because of the large
amount of expansion in the radiation era, perturbations in the
cyclic model do not build up on a given comoving scale, 
even without much expansion from dark energy. 
%(It was originally
%thought that a long dark energy phase every cycle would be necessary
%to keep diluting perturbations away.)  
Thus, our assumption in Sec.~\ref{sec:quantum}, that there is a long-lasting 
dark energy phase, is not necessary.

\section{Global Structure of the Cyclic Universe}
\label{sec:global}

Perturbations generated during the dark energy phase start off with an 
amplitude of order the Hubble radius during the dark energy phase,
say $10^{-60}$.  They are amplified during ekpyrotic contraction.
After passing through the bounce they must have an amplitude of order
$10^{-5}$ in order to match observations.  Hence there must have been a net
amplification of the order of $10^{55}$ on the largest scales!  Nothing in this 
argument 
for the amplification is unique to quantum-generated perturbations:
classical perturbations are amplified as well.  Since there are no
dynamical effects able to suppress power on scales that never enter the
horizon, we are forced to conclude that perturbations generated two or
more cycles ago should today have an amplitude formally far in excess
of unity.  In this section, we  investigate these
exceedingly long wavelength perturbations in more detail
and discuss their 
implications for the global structure of the cyclic universe.

In order to track down this large amplification, we first need to
understand how a general ultralocal Newtonian potential
perturbation evolves in time.  To start with we sketch a very general
derivation~\cite{gt} of the two linearly independent
solutions to the ultralocal perturbation equations.  This
derivation also provides a clue to the interpretation of such
perturbations.

One takes the unperturbed Friedmann-Robertson-Walker (FRW) metric, 
Eq.~\eqn{frw}, and considers a
(small) coordinate transformation.  This changes the metric  
according to the Lie derivative.  One then demands that the coordinate
transformation is such that the new metric takes the Newtonian gauge
form, Eq.~\eqn{perturbed}.  This restricts the form of the coordinate
transformation allowed, and it turns out that the most general form of
$\Phi$ so induced is:
\ba
\Phi= A \(1-\frac{H}{a} \int_0^t dt' a(t') \) + B
\frac{H}{a},
\labeq{phieff}
\ea
where $A$ and $B$ are constants.  
%\footnote{One might be puzzled by this
%  argument that in effect shows 
%that one can obtain different Newtonian potentials via a gauge
%transformation, bearing in mind that Newtonian gauge is usually said
%to be ``gauge-fixed''.  The resolution of the paradox is that our
%gauge transformation involves a dilation and so lies outside the scope
%of those transformations usually considered in perturbation theory,
%which preserve the Fourier basis for the modes.}
Furthermore, the corresponding
induced stress-energy tensor is adiabatic and has no anisotropic stress.
Now, we allow 
$A$ and $B$ to become slowly-varying functions of
position.  To the extent that second-order spatial gradients can be
neglected, which is the root of the ultralocal approximation, and
that the universe is adiabatic and free from aniostropic stress, we
now have the general solution for the Newtonian potential on long
wavelengths.  

Thus every ultralocal perturbation mode in a cycle can be written
as $\alpha_I 
\Phi^I$ ($I=1,2$), with the basis functions taken from
Eq.~\eqn{phieff} to be: 
\ba \Phi^1 &=& 1-\frac{H}{a} \int_0^t dt' a(t') \\
\Phi^2 &=& N_2 \frac{H}{a}.  
\labeq{bangbasis} 
\ea In
$\Phi^1$ we integrate forward to the time $t$ from the big bang at time
zero.  $N_2$ is a dimensionful normalizing factor, required since
$\Phi$ is dimensionless, and is helpfully chosen to make $\Phi^2$
unity at the time when $\phi=\phi_\text{end}$ after the bang.
$\Phi^1$ is the ``growing'' mode and $\Phi^2$ is the
decaying mode after the bang.

Approaching the crunch, it is useful to pick a new linear combination
of $\Phi^1$ and $\Phi^2$ as basis functions, namely:
\ba
\tilde{\Phi}^{1} &=& 1+ \frac{H}{a} \int_t^{t_\text{cr}} dt' a(t') \\
\tilde{\Phi}^2 &=& \tilde{N}_2 \frac{H}{a}.
\ea
Here $t_\text{cr}$ is the time of the crunch, and $\tilde{N}_2$ is a
different normalizing factor to $N_2$, now chosen to make
$\tilde{\Phi}^2$ unity at the time when $\phi=\phi_\text{end}$ again,
on the way to the crunch.  $\tilde{\Phi}^1$ is the ``decaying'' mode
and $\tilde{\Phi}^2$ is the growing mode going in to the
crunch.  Note that we have used these formulae in deriving the behaviour 
  of the modes in the kinetic phases near to the bounce in
  Sec.~\ref{sec:matching}, and in getting the approximation $1/t$ for 
  for the growing mode during the ekpryrotic phase in
  Sec.~\ref{sec:potential}.

Our perturbation $\alpha_I \Phi^I$ may be equivalently rewritten as
$\tilde{\alpha}_I \tilde{\Phi}^I$.  Knowing how the two sets of basis
functions are related, the two sets of expansion coefficients are
related via the matrix equation 
$\tilde{\bm{{\alpha}}}=\bm{N}\bm{{\alpha}}$
, with the matrix $\bm{N}$ of the form:
\ba
\bm{N}=
\begin{pmatrix} 1& 0 \\ n & -\varepsilon \end{pmatrix}, \labeq{ndef}
\ea
where $n \equiv 
- \int_0^{t_\text{cr}}  dt' a(t')/ \tilde{N}_2 $ and $\varepsilon \equiv
N_2 / \tilde{N}_2$.  For a typical cycle, $n$ is very large ($\,\sim
a_\text{max} t_\text{cr} / (a_\text{max} / H(\phi_\text{end})) \sim
\sqrt{-V_\text{end}/ V_0} = e^{N_\text{ekp}}\, $) and
$\varepsilon$ is very small ($\sim
e^{-N_\text{rad}-N_\text{dark}-2 \gamma/3}$).   

As discussed earlier in Sec.~\ref{sec:matching}, it is very helpful to
decompose a perturbation into growing and 
decaying modes near the crunch for the purposes of matching it through
the bounce.  We now have $\tilde{\Phi}^1$ and $\tilde{\Phi}^2$ ready for
this purpose before the bounce.  
The algebraic form for $\Phi^1$ and $\Phi^2$ will again
serve admirably for giving the growing and decaying modes after the
bounce, with $t$ now measured from this bounce and $N_2$ redefined for
this passing of $\phi_\text{end}$.  We write these new modes for the
next cycle as $\Phi_\text{next}^I$, and our perturbation after passing
through the bounce will be written as $\alpha^\text{next}_I
\Phi_\text{next}^I$.  We can now give the explicit form for the
mode-matching matrix using 
the work of Tolley, Turok
and Steinhardt, as promised earlier in Sec.~\ref{sec:matching}.
With $\bm{\alpha}^\text{next}= \bm{M}_\text{TTS}
 \, \tilde{\bm{\alpha}}$, then:
\ba
\bm{M}_\text{TTS}= \begin{pmatrix} -1 & m \\ 0 & 1 \end{pmatrix}.
\labeq{mtts}
\ea    
Here $m = v^2  e^{\sqrt{8/3}
   \, \phi_\text{end}}$ where $v$ is the (non-relativistic)
relative speed of the branes at collision.  For a typical model $m$ is
small but not particularly so.  

This analysis allows us to follow an ultralocal perturbation
forward from one cycle to the next; if it is described by the
coefficients $\bm{\alpha}$ in the one, it will be described in terms
of equivalent mode functions by the coefficients
$\bm{\alpha}^\text{next}$ in the next, with:
\ba
\bm{\alpha}^\text{next}=\bm{M}_\text{TTS} \bm{N} \bm{\alpha}.
\ea
Now, we can find the most positive eigenvalue of the combined matrix
$\bm{M}_\text{TTS} \bm{N}$ and thus finally extract the the
amplification factor per cycle.  With typical values for $m$, $n$ and
$\varepsilon$ as indicated above, this eigenvalue turns out to be
approximately $mn$ and is indeed large because $m$ is so large.  We have
thus confirmed the 
heuristic argument given at the start of this section for a large
amplification of ultralocal perturbations from cycle to cycle.
Requiring this amplification be enough to take quantum fluctuations
to $10^{-5}$ is
one of the conditions on $T_\text{rh}$ considered in
Ref.~\cite{Khoury:2003rt}, as mentioned in Sec.~\ref{sec:potential}.

What are we to make of this amplification for classical perturbations?
It certainly seems that a 
global view of 
a universe cycling everywhere with only small perturbations must break
down after a couple of bounces.  On the other hand, we have seen in
previous sections that as far as physical observers with their
cosmological horizons are concerned, the cycling can continue
indefinitely.  We believe the correct interpretation is that, as time
passes by, widely separated parts of the 
universe begin to cycle independently of one another, which precludes a
global FRW picture for the entire universe.  Nonetheless, in any given
observer's horizon, the universe appears to be FRW with
perturbations small enough that this region is able to continue
cycling. 

Perturbations on scales
larger than one Hubble horizon still have a small amplitude before
the onset of ekpyrotic 
contraction. However, they
simply correspond to a small change in that observer's background FRW
model. So by ``recalibrating'' the background model before ekpyrotic
contraction, all superhorizon perturbations can be removed.  Power on
subhorizon scales will be practically unaffected by this change.  This
``recalibration'' might lead to small amounts of space curvature in
the new FRW background, but this has negligible effect on the cyclic
history~\cite{Gratton:2003pe}.    

To show that ultralocal perturbations around some point just
correspond to a change in the background model, we effectively invert
our above derivation of the ultralocal behavior of the Newtonian
potential.  Around a chosen point, both the value and first spatial
derivatives of the real-space Newtonian potential can be set to zero with a
dilatative gauge transformation.  Furthermore, the anisotropic second
spatial derivatives can also be removed, leaving one as claimed in a
different FRW universe with perhaps some modest amount of spatial curvature
corresponding to the isotropic second spatial derivatives. 

Since different patches suffer different dilatative gauge
transformations and then have independent Fourier-expanded
perturbations, it is clear that we should not expect to be able to sew
them back together again at later times and recreate a single global
FRW solution with small Fourier perturbations; widely-separated parts
of the universe are cycling independently and out of synch.

\section{Homogeneity, Isotropy and  Flatness without Inflation}
\label{sec:flatnesspuzzle}

The previous sections have emphasized the importance of the ekpyrotic
phase and the  
overall expansion of the universe over the course of  a cycle (as illustrated
in \fig{timeline})  in  
understanding the behavior of perturbations. We now  
examine their role in explaining why the universe is so homogeneous, isotropic  
and flat today.  
 
When the cyclic model was originally introduced~\cite{Steinhardt:2002ih}, it was 
thought that the dark energy phase played the critical  
role in making the universe homogeneous, isotropic and flat,  
as well in ensuring that the cyclic solution was a stable  
attractor. This being the case, some would argue that the  
cyclic model should rightly be regarded as a variant of the  
standard inflationary scenario, since the dark energy phase  
can be viewed as a period of very low energy inflation.  
However, as the cyclic model has become better understood, we  
have learned that the dark energy phase plays only a  
supplementary role in smoothing and flattening the universe.  
In fact, as we shall now explain, homogeneity, isotropy and flatness can all be 
achieved even without the dark energy phase, as  
was first suggested by the ``cosmic no-hair theorem" proved in  
Ref.~\cite{Erickson:2004}.

First, it is already clear from Secs.~V--VII that dark energy is not
needed to make the universe homogeneous.  
If it were, 
we would have had  to impose the condition that
$N_\text{dark} \gg 1$, analogous to the condition 
that the number of e-folds of inflation must satisfy
$N_\text{inflation} \gtrsim 50$.  
In actuality, we explicitly assumed 
 $N_\text{dark} = {\cal O}(1)$ and showed that the universe is homogeneous
 after each bang when one takes into account the slowly
 contracting ekpyrotic phase.

For the curvature, we need to track what happens 
to $\Omega_\text{K} \propto 1/(a H)^2$.  
During the ekpyrotic phase with $w \gg 1$, $a$ shrinks by a small amount $e^{-
2N_\text{ekp}/(3(1+w))}$ but $H$ grows by a huge factor,  $e^{N_\text{ekp}}$; so 
the net effect is that $\Omega_\text{K}$ is suppressed by a factor of roughly  
$e^{2 
N_\text{ekp}}$.  During the contracting,
kinetic energy dominated phase with $w \approx 1$, and the subsequent expanding 
kinetic and radiation-matter dominated phases, $a H$ undergoes a net shrinkage  
by a factor of $e^{ 
 N_\text{rad}+4 \gamma/3}$, as shown in Fig.~4, so $\Omega_\text{K}$ is now 
enhanced by a 
factor of $e^{2 N_\text{rad}+ 8 \gamma/3}$.  From our key result
$N_\text{ekp} \approx  2 (N_\text{rad} + \gamma)$, though, one finds that the 
suppression of 
curvature during the contracting phase far exceeds the enhancement during the 
expanding phase, resulting in a net suppression by a factor of 
$e^{-(2 N_\text{rad}+4 \gamma/3)}$,  
 a huge net suppression of the curvature even for 
$N_\text{dark}=0$.  
This suppression repeats every time the universe goes through an 
ekpyrotic phase.

For the anisotropy, a similar analysis applies.  The anisotropic universe can be 
described 
by the Kasner metric. The anisotropy in a  Kasner universe is characterized 
by a term in 
the Friedmann equation proportional to $a^{-6}$.  The energy density in $\phi$, 
though, 
grows much faster during the ekpyrotic phase, as 
$a ^{-3 (1+w)}$.   If $H$ grows as $e^{N_\text{ekp}}$, 
then the scale factor $a$ shrinks by 
a factor of $e^{-2N_\text{ekp}/(3(1+w))}$.  Hence, the ratio of anisotropy to 
the scalar 
field energy density shrinks by a net factor of $e^{-2N_\text{ekp}(3w -
1)/(3(1+w))}$ or 
roughly $e^{-2N_\text{ekp}}$ in the limit $w \gg1$.  
During the kinetic contracting and expanding phases, 
 the ratio is fixed.  During the radiation and matter dominated phases,
 the anisotropy is only further suppressed.  
Hence, like the inhomogeneity and curvature, the anisotropy undergoes a net  
exponential suppression during each and every cycle.

Recall that, in standard big bang cosmology, a puzzling aspect of the 
large-scale homogeneity and isotropy of the universe is 
that 
distant regions within the observable horizon
were not causally connected in the past.  
 Inflation addresses this aspect by rapidly stretching a tiny, causally 
connected region 
by a huge exponential factor.  In the cyclic model, causality is 
not an issue in the first place
because the region that evolved to form the observable
universe today was  only a few meters or kilometers 
across during the previous cycle, easily 
small enough to have been in causal contact with itself during the previous
radiation and matter dominated phases.    However, in principle,
the 
universe could be causally connected and still not be 
homogeneous, isotropic and flat on large scales.  What we have shown in this 
paper is that the ekpyrotic phase and, in particular, the relation 
$N_\text{ekp} \approx  2 (N_\text{rad} + \gamma)$ automatically insures 
that it is.

\section{Conclusions}
\label{sec:conclusions} 

In this paper, we have examined the generation and evolution of 
perturbations over many cycles.
First, we have shown that the ekpyrotic phase suffices
to make the universe smooth, isotropic and flat on large scales.
As for the perturbations, we have explained how
the galaxies and large scale 
structure in any given cycle are generated by  
the quantum fluctuations in the preceding cycle without interference from 
perturbations or structure generated in earlier cycles and without interfering
with structure generated in later cycles.   Furthermore, we have examined
the global structure of the cyclic universe.  
Although the universe can be described as a nearly uniform 
Friedmann-Robertson-Walker within any observer's horizon, 
we find that global structure is  
more complex and cannot be characterized
by a uniform Friedmann-Robertson-Walker 
picture.
  
An important corollary of our results is that neither an extended dark energy 
phase 
nor any other form of inflation
is  needed to solve the horizon and flatness problems.  
Instead, 
the universe is made sufficiently smooth, isotropic and flat during each
ekpyrotic phase in which the universe contracts with $w \gg 1$.  
Hence, the cyclic model should not be construed as a variant of inflation.  
Rather,
the ekpyrotic contraction mechanism should be viewed as a genuinely novel 
approach for solving the classic cosmological problems.   

A fuller understanding of the bounce and a fundamental derivation of
$V(\phi)$  remain the most pressing issues for the cyclic model.
Major surprises there aside, our work has shown that the cyclic model
is in good shape as a candidate for a complete cosmology for the universe.

\section*{\bf APPENDIX: THE  $w\gg1$ EXPANDING PHASE}

In the cyclic model, the scalar field has to  acquire 
a boost to its kinetic energy at or after every brane collision,
in order to overcome the additional Hubble damping due to 
the radiation and to make it back onto
the potential plateau. 

As discussed in the original papers  
\cite{Steinhardt:2002ih,Steinhardt:2001st},
the
boost may be parameterized as
\be 
\left.{d e^{\sqrt{3/2} \phi} \over d t}\right|_\text{out} = -
(1+\chi) \left.{d e^{\sqrt{3/2} \phi} \over d t}\right|_\text{in}, 
\labeq{h5eq}
\ee
with $\chi$ a small parameter. 
Such a boost can be produced either by the production
of extra radiation on the negative tension brane, or by the nonminimal coupling
of $\phi$ to matter, which drives it positive in the expanding phase.

Both on the way in to the bounce on the way out from it, the 
energy density of the universe is dominated by the kinetic energy
of the scalar field. For small $\chi$, the outgoing solution is
nearly the time reverse of the incoming one and, 
as the field $\phi$ crosses $\phi_\text{end}$, the solution is close
to the time reverse of the scaling solution, Eq.~\ref{scalingsol}.
However, the scaling solution is not an attractor in the expanding
phase, so small deviations from it grow with time. The modest
increase in scalar field kinetic energy, parameterized by $\chi$,
causes the 
kinetic energy of $\phi$ to eventually overwhelm the potential
energy, so that the solution enters a second kinetic energy dominated phase. 

We compute the value of $\phi$ where this second expanding kinetic
phase begins by eliminating $t$ in favor of $\phi$  in the background equations 
\eqn{phieq} and \eqn{heq}, obtaining
\be
{d H \over d \phi} = - \sqrt{3 H^2-V \over 2}.
\labeq{newheq}
\ee
Approximating $V \approx -V_0 e^{-c \phi}$, we then change variables
to $h\equiv H e^{c \phi/2}$ to remove the leading dependence
in the scaling solution, getting
\be
{d h \over d\phi} = {c\over 2} h - \sqrt{3 h^2 + V_0 \over 2},
\labeq{hheq}
\ee
from which the fixed point scaling solution $h_\text{sc} =
\sqrt{V_0/({1\over 2} c^2-3)}$ 
is recovered. Now, we can describe the effect of the small perturbation $\chi$
by linearizing \eqn{hheq} about the scaling solution, obtaining
\be
{d \delta h \over d\phi} = \left({c\over 2}-{3\over c}\right) \delta h.
\labeq{perthheq}
\ee
Thus, the small perturbation $\delta h$ 
grows exponentially with $\phi$. The initial conditions
for $\delta h$ are found from the scaling solution, 
$H^2 = \dot{\phi}^2 /c^2$, 
the Friedmann equation and 
the definition of $h$ to be $\delta h/h_\text{sc} = (c^2/6)
(\delta \dot{\phi} /\dot{\phi})= c^2 \chi /6$. The perturbation grows
until $\delta h /h_\text{sc}$ is of order unity, when
\be
\phi - \phi_\text{end} \approx {2\over c(1-6 c^{-2})} 
{\rm ln} {6 \over c^2 \chi},
\labeq{departure}
\ee
after which 
the potential $V(\phi)$ becomes irrelevant, and, from \eqn{newheq} 
or \eqn{hheq},
$H\propto e^{-{\sqrt{3/2}} \phi}$, the expanding kinetic energy 
dominated solution. 
If $c$ is large and $\chi$ is not extremely small, 
the second kinetic phase starts rather soon after $\phi$ 
passes $\phi_\text{end}$. It follows that 
the $w\gg 1$ expanding phase is brief 
and can for most purposes be safely ignored.

\begin{acknowledgments}
  
  We thank Martin Bucher, Justin Khoury, Andrew Liddle, Jo\~{a}o
  Magueijo, Patrick McDonald, Burt Ovrut and Jim Peebles for useful
  discussions.

This work was supported in part by NSERC of Canada (JKE), by US
Department of Energy grant DE-FG02-91ER40671 (SG and PJS), and PPARC
(SG and NT). SG acknowledges the hospitality of the Relativity and
Gravitation Group in DAMTP while this work was being completed.

\end{acknowledgments}

\bibliography{cosmic16.bib}

\end{document}